# Tailoring Relaxation Time Spectrum in Soft Glassy Materials

Manish Kaushal, Yogesh M. Joshi[*]

Department of Chemical Engineering,

Indian Institute of Technology Kanpur, Kanpur 208016 INDIA

* Corresponding Author, E- mail: joshi@iitk.ac.in, Tel.: +91-512-2597993, Fax: +91-512-2590104

Abstract

Physical properties of out of equilibrium soft materials depend on time as well as deformation history. In this work we propose to transform this major shortcoming into gain by applying controlled deformation field to tailor the rheological properties. We take advantage of the fact that deformation field of a certain magnitude can prevent particles in an aging soft glassy material from occupying energy wells up to a certain depth, thereby populating only the deeper wells. We employ two soft glassy materials with dissimilar microstructures and demonstrate that increase in strength of deformation field while aging leads to narrowing of spectrum of relaxation times. We believe that, in principle, this philosophy can be universally applied to different kinds of glassy materials by changing nature and strength of impetus.

# I. Introduction

The equilibrium materials, such as polymer solutions and melts, dilute emulsions and suspensions, wormlike micelles, etc., when perturbed away from the equilibrium state, explore their phase space over the finite time scales. For such materials the relation between any impetus (for example stress field, temperature gradient, etc.) and the consequent response (such as strain or strain rate, heat flow, etc.) is unique.[1] This relation, also known as 'constitutive relation,' depends only on the nature of the material and is invariant of time (or history) and any external perturbation for the equilibrium materials. On the other hand, soft materials such as concentrated suspensions and emulsions, thixotropic pastes, colloidal gels, foams, etc., owing to the constrained mobility, cannot visit their phase space over the practical time-scales, and therefore, are out of thermodynamic equilibrium.[2,3] These materials are also called glassy. Under such conditions, the material explores its intricate energy landscape as a function of time and attaints progressively lower energy sections of the same by undergoing structural organization.[4] This induces time dependent evolution of the physical properties.[5-9] Application of deformation field causes reorganization of the structure, and can take the system to those sections of landscape, which are otherwise inaccessible in absence of deformation field.[10-12] Owing to this complex time and deformation field dependence of the structure, the constitutive relation in such materials is not unique. This major shortcoming of the glassy soft materials poses a huge challenge while analyzing their physical behavior and processing the same to form useful products. In this work we propose that this very limitation of the glassy materials can be used as an advantage. We demonstrate that by controlling deformation field it is possible to tailor the rheological properties, particularly the spectrum of relaxation times, of the same. We feel that this philosophy can significantly influence an academic outlook and processing methodology of this class of materials.

Any material, subsequent to application of impetus, dissipates energy over a range of time scales. This spectrum of relaxation times associated with the material is related to distribution of energy well depths in which its constituents are trapped.[13] In a physical aging process these constituents undergo activated dynamics and attain deeper energy well depths causing the whole relaxation time spectrum to evolve to greater values.[14] This makes the relaxation dynamics progressively sluggish as a function of time.[13,14] Application of deformation field enhances potential energy of the trapped entities.[13] Owing to distribution of energy well depths, enhancement in the potential energy facilitates diffusion of entities residing in the shallower wells, while only weakly affecting those residing in the deeper wells.[15] This interplay between time and deformation field severely affects the energy well depth distribution.[14] Consequently the spectrum of relaxation times gets profoundly altered, which is known to be responsible for many fascinating and perplexing observations such as overaging,[16,17] viscosity bifurcation,[18,19] shear banding,[20,21] delayed solidification,[15] delayed yielding,[14,22] etc. in the soft glassy materials.

## II.     Materials, Sample Preparation and Experimental Protocol

In this work we have used two types of soft glassy materials. The first system is 3.5 weight % aqueous suspension of Laponite RD. Laponite was procured from Southern Clay Products Inc. In a typical procedure, white powder of Laponite was dried in oven maintained at 120℃ for 4 hours before mixing the same with ultrapure water maintained at pH 10 using ultra turrex drive for 30 min. After preparation, the suspension was stored in sealed polypropylene bottles for a period of three months. The detailed procedure of preparing Laponite suspension in aqueous media can be found elsewhere.[23]

The second system is 7 weight % suspension of fumed silica in ethylene glycol (EG). The primary particle size of fumed silica (Silicon dioxide amorphous powder) is 7nm and was obtained from Sigma Aldrich. The fumed Silica particles were mixed with EG (Loba Chemie) and stirred for 15 minutes using mechanical stirrer to make sure that all the lumps of silica particles were completely broken, and finally a homogenous suspension was obtained. This suspension was centrifuged for 5 minutes at 1500rpm. Subsequently supernatant liquid (EG) was removed and the concentrated suspension was preserved for experiments.

The rheological experiments were performed using Anton Paar MCR 501 rheometer. In this work we have used concentric cylinder geometry with inside diameter of 5 mm and gap of 0.2 mm. In a typical procedure, the samples were shear melted (mechanical quenching) under large amplitude oscillatory stress to erase the deformation history. Subsequently samples were subjected to strain or stress controlled oscillatory field having different magnitudes and 0.1 Hz frequency. In a strain controlled aging step, aging was performed by applying $\gamma = \gamma_0 \sin \omega t$ and by varying $\gamma_0$, while in case of stress controlled aging step, oscillatory field represented by: $\sigma = \sigma_0 \sin \omega t$ was applied for different values of $\sigma_0$. We call this step as controlled aging step. During this step, irrespective of the nature of deformation field (stress or strain controlled), elastic modulus shows enhancement as a function of time for both the systems. The samples were allowed to age in controlled aging step until a predetermined value of elastic modulus is reached. Subsequent to controlled aging samples were subjected to constant stress (creep) followed by recovery. All the experiments were carried out at 25°C.

We also investigate behavior of elastic and viscous modulus of both the systems as a function of frequency. However carrying out frequency sweep experiments on the time dependent materials is not straightforward as properties change over the course of experiment leading to deceptive results. We therefore performed oscillatory experiments at different frequencies as a function of time. The value of elastic and viscous modulus at a particular time leads to frequency dependence of the same.

The materials studied in this work are thermodynamically out of equilibrium materials. Under quiescent conditions these materials undergo structural evolution to lower their free energies, which manifests in enhancement in elastic and viscous modulus as a function of time. Even though the generic signatures of aging in these two materials are same, the microstructural change that causes increase in elastic modulus is very specific to the respective systems. In case of aqueous suspension of Laponite RD, the dispersed particles have disk like shape (diameter 30 nm and thickness 1 nm). In addition, in an aqueous media faces of Laponite particles have permanent negative charge, while edges of Laponite particle have weak positive charge at pH 10.[23] Owing to dissimilar charge distribution and anisotropic shape, Laponite particles share three types of interactions among each other: Repulsion between faces, Attraction between edges and faces, and van der Waals attraction among particles. Although, origin of time dependency and soft glassy behavior in aqueous suspension of Laponite is an active area of research, both attractive as well as repulsive interactions among the particles are considered to be responsible for the same.[23,24] The second system used in this work is suspension of fumed silica in Ethylene Glycol. According to Raghavan et al.[25] silanol groups (Si–OH) present on the silica particles form hydrogen bonding interactions within the continuous phase of ethylene glycol producing a three dimensional gel network. The time evolution of rheological properties in this class of materials has been attributed to strengthening of the bond strength as a function of time.[26]

**III. Results and Discussion**

In figure 1 we plot frequency dependence of elastic and viscous modulus at certain time elapsed since shear melting. It can be seen that at this time, elastic modulus is greater than viscous modulus over the range of explored frequencies. In addition elastic modulus for both the materials can be seen to be a weak function of frequency, which is a typical signature behavior of soft glassy materials.

We now discuss the results of controlled aging experiment wherein different magnitudes of oscillatory stress and strain were applied to the shear melted samples. In figures 2(a) and 3(a) we plot evolution of elastic modulus as a function of time under application of different strain amplitudes for aqueous suspension of Laponite and fumed silica suspension respectively. On the other hand, in figures 4(a) and 5(a) we plot evolution of elastic modulus under application stress controlled oscillatory field respectively for Laponite suspension and fumed silica suspension. In all these experiments we confirmed that the first harmonic in stress/strain dominated the third harmonic justifying representation in terms of $G'$. In both the samples, under the stress and strain controlled aging experiments, $G'$ is observed to increase as a function of time. The two soft glassy materials used in this work have different microstructures as discussed above. The driving force behind microstructural evolution in any glassy material is to achieve a progressively lower free energy state, which causes increase in $G'$. However it is observed that the microstructural evolution (and increase in $G'$) does not stop over practically observable timescales. We therefore allow the samples to age until $G'$ evolved to a predetermined level. Interestingly, even though every sample within a system has identical value of $G'$ at this state, it has acquired the same via different paths. It is known that the linear viscoelastic behavior of a material can be completely characterized by spectrum of relaxation times associated with the same.[27] Therefore, the very fact that various samples of this thermodynamically out of equilibrium material have acquired the same value of $G'$ by evolving through different paths suggests that the spectrum of relaxation times associated with the same may be different.

In order to quantify the breadth of relaxation time spectrum we perform creep-recovery experiments on the samples that have evolved to the same $G'$ through different paths. The corresponding creep recovery behavior subsequent to strain controlled aging for aqueous Laponite suspension is shown in figure 2(b) while that of for fumed silica suspension is shown in figure 3(b). The creep recovery behavior following stress controlled aging for aqueous Laponite suspension and fumed silica suspension is shown in figures 4(b) and 5(b) respectively. The figures 2 to 5 show that strain recovery is practically complete within 5 s, so that the total creep - recovery time is significantly smaller than the age of the samples in the controlled aging step. We also measured the yield stress (in dynamic mode) just after the controlled aging step and observe that the creep stress used for each material is of the same order as that of the yield stress. (it should be kept in mind that it is difficult to measure the yield stress of a time dependent material due to change in its properties during the experiment, therefore the estimated yield stress is merely indicative in nature).

In figure 6 we plot ultimate recovery normalized by elastic modulus and creep stress ($\gamma_\infty G'/\sigma_c$) as a function of $\gamma_0$ imposed during controlled aging for both the soft glassy materials for the data shown in figure 2 and 3. Interestingly, it can be seen that with increase in $\gamma_0$, a consistent decrease in $\gamma_\infty G'/\sigma_c$ is observed. In figure 7 we plot normalized ultimate recovery ($\gamma_\infty G'/\sigma_c$) as a function of magnitude of normalized oscillatory stress ($\sigma_c/G'$). Interestingly, similar to that observed for strain controlled aging, $\gamma_\infty G'/\sigma_c$ for both soft glassy systems is also observed to decrease with increase in $\sigma_0$. From figures 3 to 5 we can also estimate instantaneous compliance ($J_i$) induced in the material subsequent to application creep stress. Similar to trend in $\gamma_\infty$, instantaneous compliance is also observed to decrease with increase in magnitude of stress/strain oscillations applied during controlled aging (not shown). In rheology literature, $J_i$ or $\gamma_\infty$ of material is closely related to spectrum of relaxation times.[27,28] Therefore the observed trend suggests consistent change in the breadth of spectrum as a function of $\gamma_0$ or $\sigma_0$ applied during controlled aging. In figures 6 and 7 we have normalized $\gamma_\infty$ with $\sigma_c/G'$. The term $\sigma_c/G'$ represents instantaneous strain induced in the material for a single mode Maxwell model as soon as creep stress is applied. Interestingly $\sigma_c/G'$ is of the order of experimentally observed value but not identical. The term $\gamma_\infty G'/\sigma_c$, therefore in principle, represents magnitude of recovered strain compared to instantaneous strain induced in the material if material were single mode with modulus $G'$.

The results of figure 6 and 7 suggest that irrespective of nature of applied oscillatory field (stress/strain controlled) during controlled aging, $\gamma_\infty$ (at same $G'$ for a given system) is observed to decrease with increase in $\gamma_0$ or $\sigma_0$. This clearly suggests that the flow field applied during controlled aging step profoundly affects the spectrum of relaxation times. In order to analyze behavior of $\gamma_\infty$ to ascertain breadth of relaxation time spectrum, we consider viscoelastic framework represented by Boltzmann superposition principle. It is important to note that various samples, which have evolved through different deformation fields (but have same $G'$) over a prolonged duration, have been subjected to creep-recovery protocol over a very brief period of time. Therefore, change in rheological properties over the duration of creep-recovery can be neglected, a good approximation often applied to glassy materials.[29] Consequently we assume that every sample is invariant of time over the duration of creep-recovery, and obeys Boltzmann superposition principle given by:[28]

$$\sigma(t) = \int_{-\infty}^{t} G(t-t')\frac{d\gamma}{dt'}dt', \tag{1}$$

where $\sigma$ is shear stress, $\gamma$ is shear strain, $t$ is present time, and $t'$ is past time. In a typical procedure, we apply creep stress over duration of $t_1$ ($\sigma = \sigma_c$), and subsequently let the material recover ($\sigma = 0$). If we consider application of creep at $t' = -t_1$, equation (1) for any time $t > 0$ can be represented as:

$$0 = \int_{-t_1}^{0} G(t-t')\dot{\gamma}(t')dt' + \int_{0}^{t} G(t-t')\dot{\gamma}(t')dt' \tag{2}$$

It can be seen in all creep plots that strain induced in the material due to application of creep can be considered to have constant slope (except the initial transient) leading to $\dot{\gamma}(t') = \dot{\gamma}_0$ (We also fitted a straight line to strain evolution data and found that value coefficient of determination $R^2$ for Laponite suspension is always above 0.95, while that of for Silica suspension is always above 0.98). This facilitates analytical solution of eq. (2) for $\gamma_\infty$ by taking Laplace transform. If we assume spectrum of relaxation times is given by $\tau_k$ (with $\tau_k > \tau_{k+1}$) and $G_k$, relaxation modulus is given by: $G(s) = \sum_k G_k e^{-s/\tau_k}$. Expression of ultimate recovery obtained from eq. (2) in terms of relaxation time distribution is given by:

$$\gamma_\infty = \int_0^\infty \dot{\gamma}(t')dt' = \sigma_c \left[\sum_k G_k \tau_k^2 \left(1 - e^{-t_1/\tau_k}\right)\right] \Big/ \left[\sum_k G_k \tau_k\right]^2. \tag{3}$$

Relationship between ultimate recovery and spectrum of relaxation times can be significantly simplified by assuming an empirical expression for relaxation time distribution proposed by Spriggs:[27,28,30] $\tau_k = \tau_0 k^{-\alpha}$ and $G_k = G_0 k^{-\alpha} / \sum_k k^{-\alpha}$, where $\alpha$ represents breadth of the spectrum of relaxation times. Greater the value of $\alpha$ is, narrower is the breadth. Spriggs distribution is popularly used in the rheology literature and has been validated experimentally for polymer solutions and melts for $\alpha$ in the range 2 and 4.[27,28] Incorporating Spriggs distribution in equation (3), we get:

$$\gamma_\infty G_0 / \sigma_c = \left[\sum_k k^{-\alpha}\right]\left[\sum_k k^{-3\alpha}\left(1 - \exp(-t_1 k^\alpha / \tau_0)\right)\right] \Big/ \left[\sum_k k^{-2\alpha}\right]^2. \tag{4}$$

We compare samples aged under different magnitudes of oscillatory strain/stress but having same elastic modulus. Expression of elastic modulus is given by: $G' = \text{Re}\left(i\omega \int_0^\infty G(t)e^{-i\omega t}dt\right)$. Incorporating expression for $G(t)$ and Spriggs spectra, we get:[31]

$$G' = G_0 \left[\sum_k \frac{k^{-3\alpha}}{\tau_0^{-2}\omega^{-2} + k^{-2\alpha}}\right] \Big/ \left(\sum_k k^{-\alpha}\right). \tag{5}$$

For soft glassy materials in solid state ($G' >> G''$), relaxation time is significantly greater than inverse of frequency: $\tau_0 \omega >> 1$, which leads to $G' \approx G_0$. Interestingly, eq. (4) shows that despite the same value of elastic modulus, ultimate recovery indeed depends on $\alpha$. In the controlled aging experiment we have employed frequency of 0.628 rad/s (0.1 Hz), which suggests dominating relaxation mode to be $\tau_0 >$ 1.6. On the other hand, creep field was applied for 30 or 40 s. Therefore, although it is difficult to quantify the precise value of $\tau_0$, we cannot rule out any possibility $t_1/\tau_0 < 1$ as well as $t_1/\tau_0 > 1$. In the limit of $t_1/\tau_0 \ll 1$, eq. (4) can be further simplified by carrying out Taylor series expansion of exponential term to yield:

$$(\gamma_\infty G_0/\sigma_c)(\tau_0/t_1) = \sum_k k^{-\alpha} / \sum_k k^{-2\alpha} . \qquad (6)$$

In Fig. 8 we plot $\gamma_\infty G_0/\sigma_c$ as a function of $\alpha$ for different values of $t_1/\tau_0$ by solving eq. (4), while in the inset we plot eq. (6) for $t_1/\tau_0 \ll 1$. It can be seen that for any finite creep time $t_1/\tau_0$, $\gamma_\infty$ decreases with increase in $\alpha$. This result suggests that smaller $\gamma_\infty$ corresponds to narrower spectrum of relaxation times.

The results of Figs. 6 and 7 clearly indicate that $\gamma_\infty$ decreases with increase in $\gamma_0$ or $\sigma_0$ applied during the controlled aging step. Therefore, comparison of experimental observation with the proposed model suggests that the shape of spectrum of relaxation times becomes narrower with increase in the strength of oscillatory deformation field. In a soft glassy rheology (SGR) framework, for strain induced in the material having magnitude $\gamma$, potential energy enhancement of particles is given by: $\tfrac{1}{2}k\gamma^2$ assuming the affine deformation [as shown in figures 2 to 5, strain induced in the sample is very small <0.3, which justifies the affine assumption].[13] Consequently particles residing in wells with depth shallower than $O(\tfrac{1}{2}k\gamma^2)$, will undergo faster rejuvenation. Such particles will get arrested in a new well whose depth will be randomly chosen from the available distribution.[13,15] However, since deformation is continuous, wells significantly deeper than $\tfrac{1}{2}k\gamma^2$ will only be available to choose from, and therefore will get progressively populated. During the controlled aging step, greater amplitude of strain induced in an oscillatory flow (which is the case of higher magnitude of stress or strain imposed during controlled aging) will therefore cause progressively narrowing of the spectrum of relaxation times as a function of time. Interestingly our experimental results on both soft glassy materials with different microstructure demonstrate this phenomenon very well, emphasizing universality associated with the same.

Usually, in order to estimate the breadth of relaxation time distribution, stress relaxation is more commonly used. The decay of stress relaxation modulus is usually fitted using KWW function given by: $G(s) = G_k \exp\left(-\left(s/\tau\right)^\beta\right)$, wherein decrease in value of exponent $\beta$ below unity represents broadening of relaxation time spectrum.[32] We also carried out stress relaxation experiments to estimate relaxation time spectrum (not shown). However, owing to soft glassy nature, relaxation of stress is extremely sluggish and takes significantly larger time than the age of the material. Since material undergoes physical aging during this extended period, it is expected to affect relaxation of stress leading complications in the analysis. Recovery of strain, on the other hand is complete significantly faster compared to relaxation of stress. We have therefore used strain recovery protocol in this work. Interestingly stretched exponential function can also be used to fit the decay of intensity autocorrelation function to estimate relaxation time and exponent $\beta$. Ruocco and coworkers[33] studied effect of strain rate on relaxation time and $\beta$ by carrying out in situ dynamic light scattering in a rheometer. They observed that for higher shear rates, stretch exponential exponent $\beta$ (associated with decay of normalized intensity autocorrelation function) increases. Increase in $\beta$ directly corresponds to narrow distribution of relaxation times.[32]

Furthermore, we believe that the methodology described in this work, although presented using a rheological example, can in principle be extended to different types of *impetus – glass* pairs (Impetuses that induce energy in the material, such as stress/strain field, electric field, magnetic field, etc. and its suitable glass counterpart such as soft, molecular or spin glasses). Similar to soft glassy rheology model argument, if application of certain magnitude of electric or magnetic field induces energy in constituents of glass affinely, then up to a certain extent, relaxation modes will not be available to the aging material and qualitatively the same effect will be observed.

## IV. Conclusion

Soft glassy materials, owing to their out of thermodynamic equilibrium nature, are known to show strong time and history dependence. In this work we propose that this apparent liability can be used as an advantage by applying controlled deformation field. Using two different soft glassy materials, we demonstrate that prolonged application of greater amplitude of stress or strain during physical aging can cause narrowing of spectrum of relaxation times. We believe that this work opens up new frontiers in designing soft glassy products with desired properties.

**Acknowledgement:** We would like to thank Department of Atomic Energy – Science Research Council (DAE-SRC) for financial support.

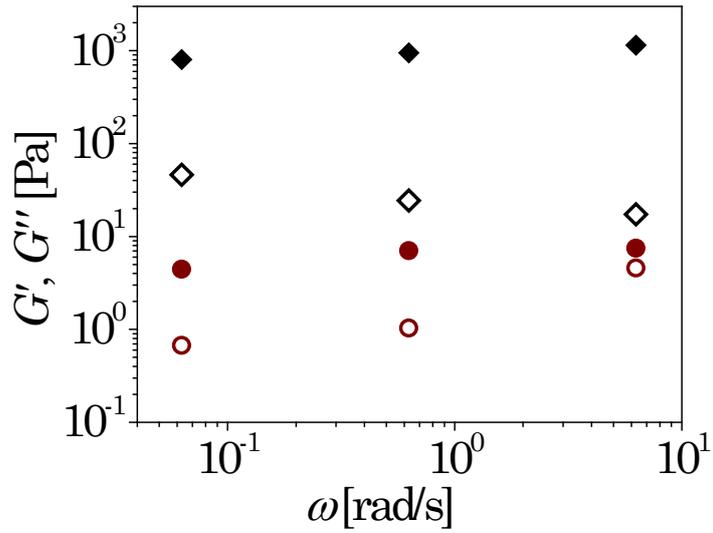

**Fig. 1.** Evolution of $G'$ (closed symbols) and $G''$ (open symbols) as a function of angular frequency for Laponite suspension (diamond, $G' \approx \omega^{0.08}$) and Silica suspension (circle, $G' \approx \omega^{0.11}$). The frequency dependence is obtained by exploring temporal evolution of $G'$ and $G''$ at different frequencies. The data mentioned in figure belongs to Laponite suspension having age of 30 min while silica suspension having 13 min.

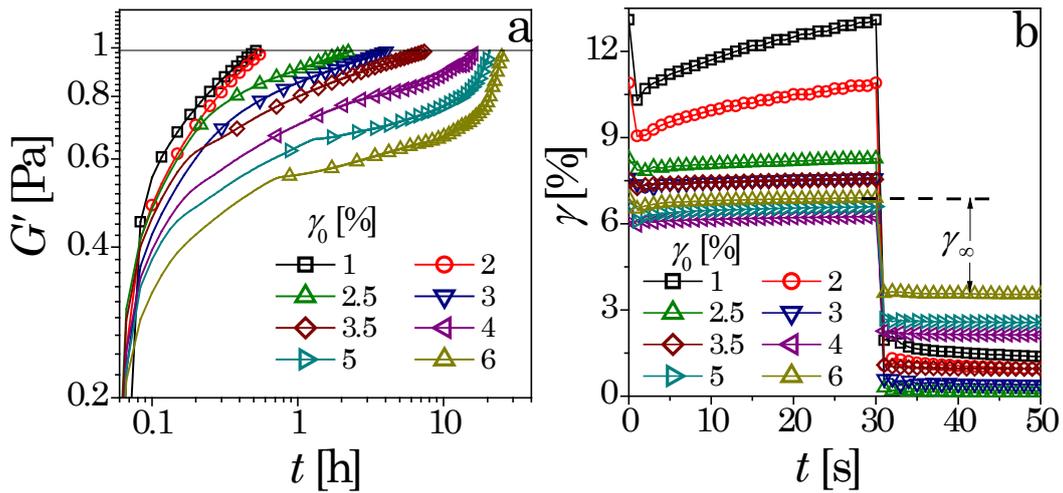

**Fig. 2.** Evolution of $G'$ during controlled aging under oscillatory strain field having different amplitudes ($\gamma_0$) for Laponite suspension is shown in (a). The subsequent creep ($\sigma_c$ =50 Pa)-recovery plots are shown in (b).

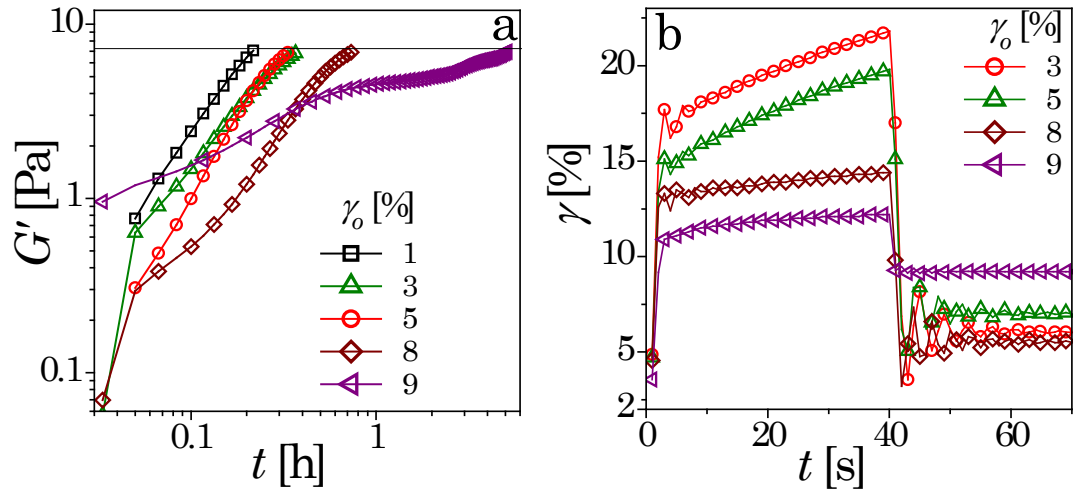

**Fig. 3.** Evolution of $G'$ during controlled aging under oscillatory strain field having different amplitudes ($\gamma_0$) is shown for Silica-EG suspension (a) along with the corresponding creep ($\sigma_c$=0.6 Pa)-recovery plots (b). In figure (b) we have not shown creep recovery curve belonging to 1 % strain during controlled aging as strain induced during creep is much larger to show the remaining creep curves properly.

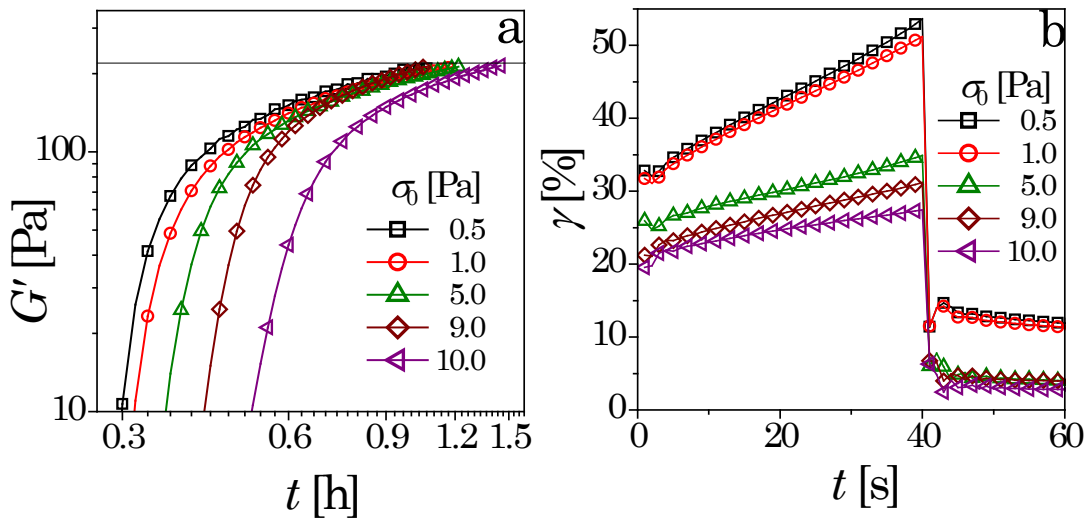

**Fig. 4.** Evolution of $G'$ is plotted as a function of time (a) during stress controlled aging for Laponite suspension. The right figure (b) shows corresponding creep ($\sigma_c$=25 Pa)-recovery plots.

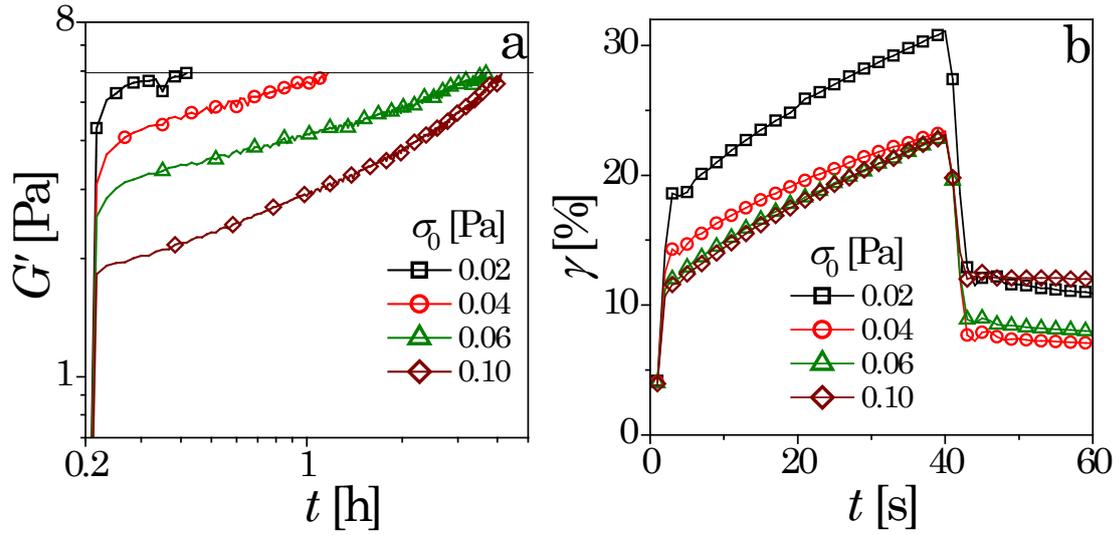

**Fig. 5.** Time evolution of $G'$ for Silica-EG suspension during stress controlled aging is plotted in (a). The subsequent creep ($\sigma_c$ =0.5 Pa)-recovery is plotted in figure (b).

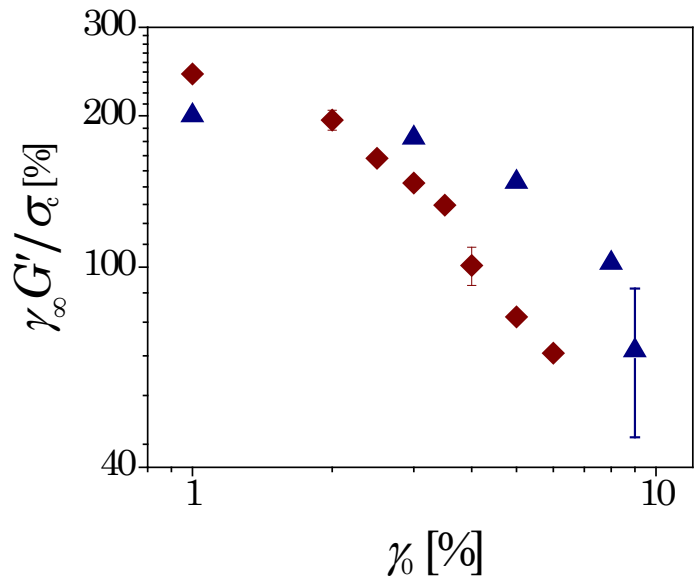

**Fig. 6.** Normalized ultimate recovery ($\gamma_\infty G'/\sigma_c$) plotted against amplitude of strain ($\gamma_0$) imposed during strain controlled aging. Wine diamonds and blue triangles represent Laponite suspension and silica EG suspension respectively.

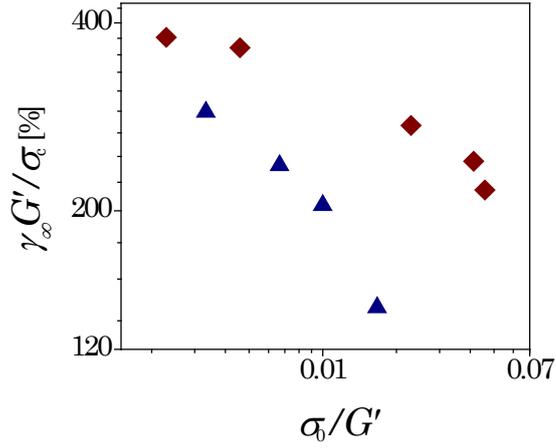

**Fig. 7.** Normalized ultimate recovery ($\gamma_\infty G'/\sigma_c$) plotted against normalized amplitude of stress ($\sigma_0/G'$) imposed during stress controlled aging. Wine diamonds and blue triangles represent Laponite suspension and silica EG suspension respectively.

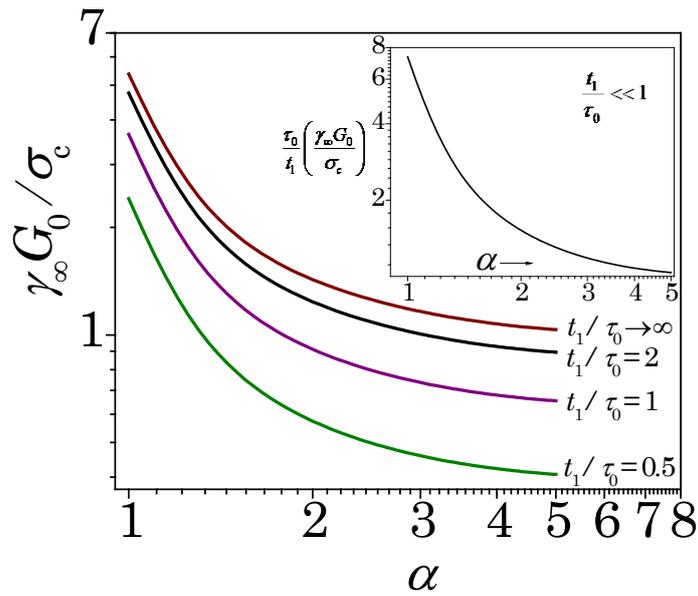

**Fig. 8.** Normalized ultimate recovery ($\gamma_\infty G_0/\sigma_c$) is plotted against $\alpha$ as given by eq. 4, for different values of $t_1/\tau_0$. In the inset $(\gamma_\infty G_0/\sigma_c)(\tau_0/t_1)$ is plotted as a function of $\alpha$ in the limit of $t_1/\tau_0 \ll 1$ as described by eq. 6.